\documentclass{article}
\usepackage[utf8]{inputenc}
\usepackage{graphicx}
\usepackage{longtable}
\usepackage{multicol}
\usepackage{multirow}
\usepackage{appendix}
\usepackage{amsmath,amssymb,amsfonts}
\usepackage{url}

\title{\Huge{Systematic Mapping Protocol} \\ \LARGE{Feature Modeling Tools} \\}
\author{Authors: Samuel Sepúlveda, Andrea Rivera\\ \\ Dpt. of Computing and Informatics\\University of La Frontera\\Temuco, Chile\\}

\date{Final version: March 2019}

\begin{document}

\maketitle
\newpage

\section{Introduction}
\label{Sec:Intro}

The customers and users need for new products and services according to high quality standards have increased in the last time. In that sense, the productive processes must be aligned with the organization and development process in order to achieving this goal.

Software product line (SPL) is an approach that can deal with those needs, increasing productivity without sacrificing quality. SPL is a set of software intensive systems that share a common set of features and are developed for a specific segment or domain using a defined process~\cite{clements2002}. There are several known benefits of the SPL use: reduction in the cycles of product development, increase in productivity by an order of magnitude, decrease in cost and a substantial improvement in the quality of products~\cite{ahmed2008,mcgregor2010}.

The domain analysis for SPL is, to the best of our knowledge, the most important stage in the development process. Here, feature models (FMs) are a domain analysis artifact used to describe all identified features. Furthermore, it is \textit{de facto} standard for managing variability~\cite{Collet2014}.

The aim of this paper is to synthesize the current state of research reported in the literature regarding the application domain, underlying model, origin, degree of empirical validation and quality of existing feature modeling tools used in SPL. The motivation is to check for improvement tendencies in the field, covering the period between 2000 and 2019. In particular, empirical validation of the tools, has been repeatedly pointed out as important deficiencies of the field. Also, we include an initial quality assessment for the different feature modeling tools that we will find.

We think this study may be of interest to both academic researchers and industry professionals who wish to get an updated view of feature modeling tools, which are their shortcomings and strengths in terms of some quality criteria. With this knowledge, they will better assess the potential benefits and risks associated with adopting each feature modeling tool. Furthermore, it could be of interest to researchers looking for gaps in research for doing additional studies on feature modeling tools for SPLs. In addition, we see this study as a continuation of what we have done on different aspects for FMs and modeling languages used in SPLs \cite{Sepulveda2012,Sepulveda2012a,Sepulveda2016}.

Therefore, this technical report presents the protocol definition for a systematic mapping study (SMS) that we will conduct to identify and assess the set of relevant papers on feature model tools.

The rest of the report is structured as follows. Section~\ref{Sec:Research} describes the research method to follow. Section~\ref{Sec:Threats} presents and discusses the main threats to validity, and the strategy to deal with them. Finally, Section~\ref{Sec:Conclusions} presents our conclusions and ongoing work.

\section{Research method}
\label{Sec:Research}

This study has been carried out according to the SMS methodology described by~\cite{kitchenham2010value}, as a methodology that aims to \textit{``identify all research related to a specific topic rather than addressing the specific questions that conventional SLRs address"}. Similarly, \cite{petersen2008systematic} indicate, \textit{``a systematic mapping is a method to build a classification scheme and structure a Software Engineering (SE) field of interest. The analysis of results focuses on frequencies of publications for categories within the scheme. Thereby, the coverage of the research field can be determined"}. 
In this study we will search for existing research related to feature modeling tools in the context of SPLs, and we have classified and analyzed them according to certain predefined criteria.

\begin{figure} [!ht]
\centering
\includegraphics[width=.55\textwidth]{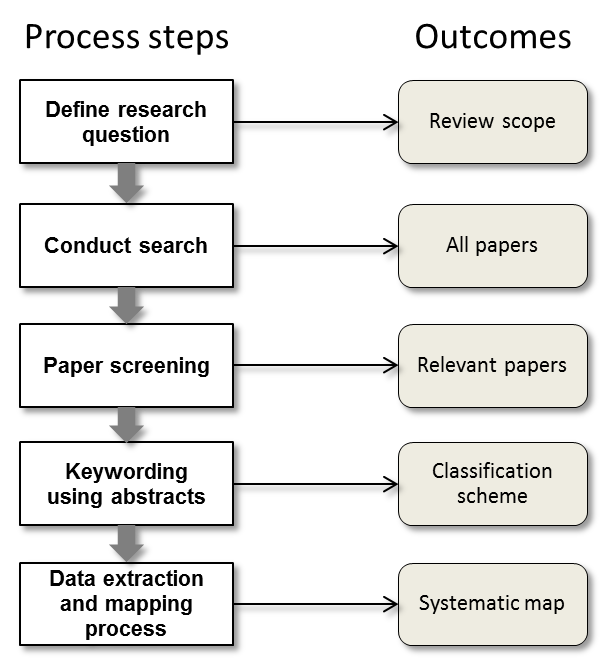}
\caption{SMS steps} \vspace{-5mm}
\label{SLRprocess}
\end{figure}

Next, in Section~\ref{ProtocolDefinition} we define the SMS protocol. Then, in Section~\ref{StudySelection_DataExtraction} we describe the study selection and in Section~\ref{PrimaryStudySelectionDataExtraction} we define the preliminary data extraction protocol.
Finally, in Section \ref{SMStoolSupport}, we briefly describe the tool support used for our SMS. The whole process followed for the SMS is shown in Figure~\ref{SLRprocess}, adapted from \cite{petersen2008systematic}.


\subsection{Protocol Definition (Planning)}
\label{ProtocolDefinition}
In this section we present the main steps performed in the protocol definition for this SMS.

\subsubsection{Aim and Need}
\label{AimAndNeed}
The aim of this SMS is twofold. On the one hand, we have established some issues about application domain, underlying feature model, origin and degree of empirical validation of feature modelling tools for SPLs. On the other hand, we have assessed the \textit{``quality level"} for the selected feature modelling tools.

Therefore, the importance of this study lies in the issues mentioned above, in addition to other aspects included in this study, namely origin of the papers, context of application, year of publication, publisher and target audience, among others. 

We think that a clear picture of all these characteristics may help professionals reduce the associate risk for choosing a tool. Additionally, we aim to foster a discussion among the members of the community about the qualities that feature modelling tools for SPLs should have in order to promote the creation and sharing of high-quality specifications.

\subsubsection{Research Questions}
\label{ResearchQuestions}
The RQs define what should be extracted from the final selected publications \cite{keele2007guidelines}. Table \ref{TablaRQ} presents the four RQs that drive our SMS, together with their contribution to the general aim.

\begin{table}[!ht]
\caption{Research Questions for the Systematic Mapping Study} 
\label{TablaRQ}

\begin{tabular}{p{0.6cm}|p{4.2cm}|p{5.8cm}}
  \hline
 \textbf{ ID} & \textbf{Question} & \textbf{Aim and Classification Schema} \\

  \hline
RQ1 & What is the feature modelling tool's application domain? & To determine if the tool is multipurpose or has been developed/used in specific domains. \\

  \hline
RQ2 & What model underlies the selected feature modelling tool? & To determine what model each tool is linked to, e.g. FODA and its variants, cardinality based model or others. \\

  \hline
RQ3 & Where have feature modelling tools for SPLs been developed? & To identify the origin of the tools: academia, industry or joint.   \\

  \hline
RQ4 & What is the degree of empirical validation of feature modelling tools in SPLs? & To examine how each selected tool was validated: with proofs of concept, through its use in industry, through case studies, through experiments, etc. \\

  \hline
\end{tabular}
\end{table}

\subsubsection{Search String}
\label{SearchString}
The search string was constructed as follows \cite{keele2007guidelines,Unterkalmsteiner2012}:
\begin{itemize}
\item From the RQs we obtained keywords.
\item From the keywords, we considered synonyms.
\item We built the search string by applying the criterion Population-Intervention-Comparison-Outcomes-Context (PICOC \cite{petticrew2005picoc}).
\end{itemize}

According to~\cite{keele2007guidelines}, \emph{population} in SE should correspond to one of the following: (1) specific SE role, (2) a category of software engineer, (3) an application area or (4) an industry group. In our case, \emph{Software Product Lines} was considered an \emph{application area}.

An \emph{intervention} in SE is defined as a methodology, tool, technology or procedure that addresses a specific issue~\cite{keele2007guidelines}. For example, performing specific tasks such as requirements specification, system testing, or software cost estimation. In our case, the intervention is part of a \emph{tool}, in particular for  \emph{Domain Engineering} stage and \emph{Feature Modelling} step.

The \emph{comparison} element is not applicable to our RQs, because they did not involve the comparison of the collected papers against any commonly used feature modelling tool or technique (the control condition).

The main \emph{outcomes} of our RQs are the origin, underlying model, application domain, together with their level of validation in the software industry.

Last, the \emph{context} represents the place where the comparison is done, for example academia, industry or both.

All different defined terms were combined with the ``AND'' boolean operator, and all the synonyms were joined to each other by using the ``OR'' operator to improve the completeness of the results. The terms, synonyms, final search string and search strategy are shown in Table \ref{TablaSearchString}.

\begin{table}[!ht]
\caption{Search String} \label{TablaSearchString}
\begin{center}
\begin{tabular}{p{2.5cm}|p{8.7cm}}
  \hline
  Terms & Feature, modelling, model software, family, product, lines, variability, tool\\

  \hline
 Combining Terms &  ``Feature modelling", ``Feature model", ``Variability", ``Software product lines", ``Tool", ``Software family", ``Product family" \\

  \hline
Search String & (``Feature modelling" OR ``Feature model" OR ``Variability") AND (`Software product lines" OR ``Product family" OR ``Software family" OR SPL) AND (``Tool")\\

  \hline
Search strategy & The string was entered sequentially into each data source, adapting it accordingly. Variations in spelling (\textit{e.g.} modelling \textit{vs.} modeling) were also accounted for.\\
 \hline
\end{tabular}
\end{center}
\end{table}

\subsubsection{Inclusion and Exclusion Criteria}
\label{InclusionExclusionCriteria}
In this study we defined both inclusion and exclusion criteria. By checking these criteria we decided whether an article was finally included or not in the SLR, based on its content.

In particular, and following the guidelines of \cite{keele2007guidelines}, grey literature (i.e. technical reports, white papers and work in progress) was excluded.

These criteria are defined in Table \ref{TablaInclusionExclusionCriteria}. 

\begin{table}[!ht]
\caption{Content-related Inclusion and Exclusion criteria} \label{TablaInclusionExclusionCriteria}
\begin{tabular}{p{2.5cm}|p{9.0cm}}
  \hline
  Inclusion criteria & Papers that address the topic of feature modelling tools for SPLs, from any of the following perspectives:
  \begin{itemize}
  \item Studies that propose feature modelling tools in software product lines.
  \item Peer-reviewed studies obtained from journals, conferences and workshops.
  \item Studies published from 2000 to 2019. 
  \item Studies published in English or Spanish.
  \end{itemize}\\
 \hline
 Exclusion criteria &  Papers that, even if they discuss proposals and tools for SPL and variability modelling, do not center specifically on feature modelling tools for SPLs:
  \begin{itemize}
  \item Studies by the same author or group of authors who do not contribute significant improvements to prior proposals in the case where there is a recent proposal.
  \item Studies not available online.
  \item Studies that deal with secondary research, such as mapping studies or systematic literature reviews.
  \end{itemize}\\
 \hline
\end{tabular}
\end{table}

\subsubsection{Protocol Validation}
\label{ProtocolValidation}

The protocol validation was performed along with the definition of each of the steps of the protocol. This validation was based on the criteria defined by~\cite{Petersen2015}, and we concretely and objectively identified how we developed our mapping study. In the appendix \ref{appendix:Evaluation} we detail the evaluation process for the SMS protocol.

The information presented in this paper corresponds to the final result (definition plus validation) of each step. 
According to the evaluation done to our systematic mapping study, we applied at least one action for each rubric criteria group established in the protocol phase~\cite{Petersen2015}. 

Considering the ratio of the number of actions
taken in our study in comparison to the total number of actions possible to be taken, the calculated ratio was 38\% (10 over 26 items).

\subsection{Primary Study Selection}
\label{StudySelection_DataExtraction}
We made a list that was as complete as possible of papers related to feature modelling tools and SPLs. This SMS dates back to 2000 and the search was conducted between March and May 2019.

\subsubsection{Search Process}
\label{SearchProcess}
We design a search strategy that consisted of an automatic search on electronic databases, eventually we consider perform a snowballing approach to complete the search.

We consider the following databases: \emph{IEEE Xplore}, \emph{ACM Digital Library}, \emph{Science Direct} and \emph{Scopus}. These sources are recognized as being among the most relevant in the Software Engineering community~\cite{keele2007guidelines,brereton2007lessons}.

\subsubsection{Pilot Selection}
Once both the inclusion and exclusion criteria and the data sources had been defined, we performed a pilot selection and extraction to ensure the reliability of the protocol.

For all the researchers involved in the selection of the primary studies, we will  verify that the manner of applying and understanding the inclusion/exclu\-sion criteria be similar for everyone (inter-rater agreement), avoiding any potential bias.

This will be tested as follows: for all the researchers individually deciding on the inclusion/exclusion of a set of papers randomly chosen from those retrieved by this pilot selection. We perform a test of concordance based on the \emph{Fleiss' Kappa statistic} as a means of validation \cite{gwet2002inter}. We consider to obtain a \emph{Kappa$\geq$0.75}, could be a value that suggests that the criteria were clear enough to apply the inclusion and exclusion criteria in a consistent way for each one of the researchers~\cite{fleiss1981statistical}.

\subsection{Preliminary Data Extraction Protocol}
\label{PrimaryStudySelectionDataExtraction}

Once both the search string and the inclusion/exclusion criteria had been tested, we launched the primary study retrieval and the data extraction phase. A summary of this phase can be seen in Table \ref{TablaProtocolReview}.

\begin{table}[!ht]
\caption{Access and Data extraction protocol}
\label{TablaProtocolReview}

\begin{tabular}{p{3cm}|p{8.5cm}}
  \hline
  Paper access & Access to each of the papers to be reviewed must be guaranteed.\\

  \hline
 Initial review of the paper & Read the title, abstract and keywords of each paper to decide the relevance to the SMS.\\

  \hline
 Review Report & Scan the whole paper and answer the following questions: 
 \begin{itemize}
 \item Why was the paper accepted/rejected? 
 \item If the paper was accepted
 	\begin{itemize}
	\item Why is the paper relevant to the SMS?
 	\item Which of the RQs does the paper answer? 
 	\end{itemize}
 \end{itemize} \\
  \hline
\end{tabular}
\end{table}

First, we will run the search string in the selected data sources, mentioned in Section~\ref{SearchProcess}. This process will turn in aprox. 1000 and 1500 results (according to pilot searchs). After that, we will eliminate the duplicates. Then, we will look through the title, abstract and keywords (if available) to get an initial impression of their thematic relevance (See Table \ref{TablaProtocolReview}). In this step the papers that will not be rejected follow on to the next step. 

Next, we will apply the format-related inclusion/exclusion criteria. We will discard papers not in English grey literature. 
In addition, we will discard papers that presented a different version of the same proposal. When the latter was the case, we retain the most current version of the proposal in the selection. 


Last, we will divide that list among the researchers, and each one apply the content-related inclusion/exclusion criteria defined in Table \ref{TablaInclusionExclusionCriteria}, obtaining the final list of selected papers. 

This information will be then jointly reviewed to collaboratively accept the final list of selected papers. The whole list, including a brief description of all the selected papers for this SMS will be summarized in a appendix on the final paper with the results of this systematic mapping.


\begin{figure}[!ht]\centering
\includegraphics[width=1.0\textwidth]{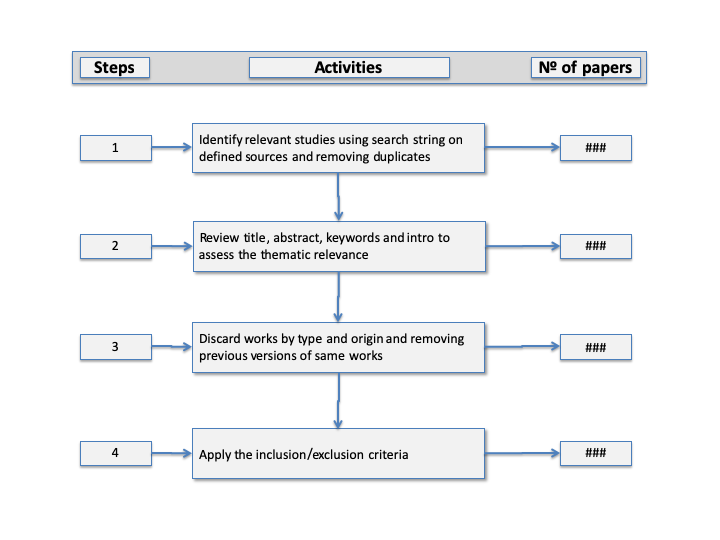}
\caption{SMS Primary Study Selection Steps.} \vspace{-1mm}
\label{stepsSLR}
\end{figure}

\subsubsection{Preliminary Data Extraction and Assessment}
\begin{figure}[!ht]\centering
\includegraphics[width=1.0\textwidth]{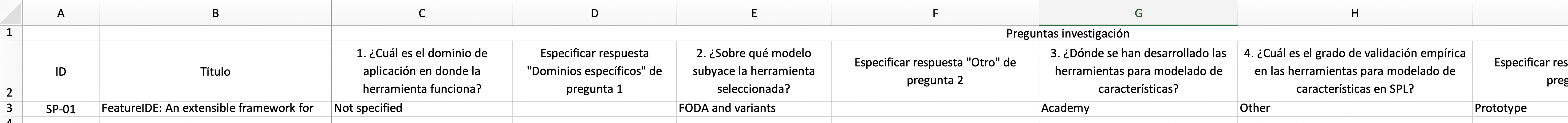}
\caption{Example for a data extraction form for the selected papers.} \vspace{-1mm}
\label{fig:paperForm}
\end{figure}

\begin{figure}[!ht]\centering
\includegraphics[width=1.0\textwidth]{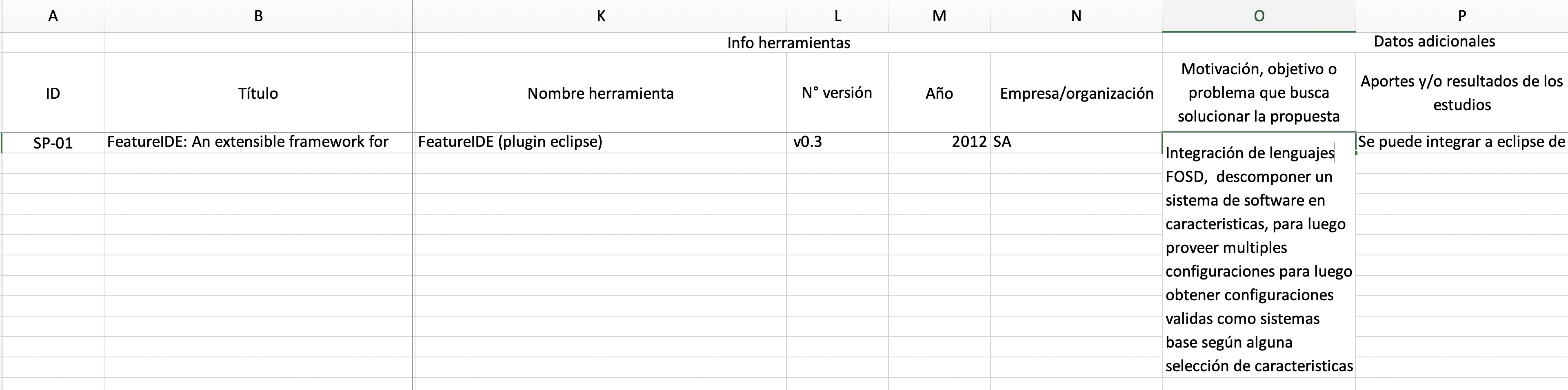}
\caption{Example for a data extraction form for the tools in selected papers.} \vspace{-1mm}
\label{fig:toolForm}
\end{figure}

For each selected paper (that will meet the inclusion criteria), we will read it, extracting relevant data in order to answer the established RQs. Figures \ref{fig:paperForm} and \ref{fig:toolForm} show en example for the data extraction form that will be used to compile the details about the paper and the tool reported.

The extracted data for each paper and their assessment strategy will be as follows: (i) Title, authors, year, (ii) Reason why the paper was initially included, (iii) Type of publication journal \textit{(SCI-JCR quartile\footnote{Journal Citation Reports, \url{ http://thomsonreuters.com/journal-citation-reports/}} or other)} or conference proceeding \textit{(CORE ranking\footnote{Computing, Research and Education, \url{ http://core.edu.au/index.php/}} or other)} and the corresponding editor, (iv) Type of  experience reported, (v) Results, (vi) Community to which the paper was directed and (vii) Tools and programming languages used. The detail of this is shown in Table \ref{TablaEXtraccionData}.



\begin{table}[!ht]
\caption{Data Extraction Protocol}
\label{TablaEXtraccionData}
\begin{tabular}{p{2.0cm}|p{8.1cm}}
  \hline
  Initial reading & The abstract, introduction, related work, conclusions and references should be read to collect background information about:
  \begin{itemize}
  \item Community (Introduction, Related Work and References)
  \item Contributions of the paper according to its authors (Abstract, Introduction and Conclusions)
  \item Possible consequences of contributions: applications, new techniques or research (Introduction, Conclusions and Future work).
  \end{itemize}\\
  \hline
  Detailed reading & The body of the article should be read in order to:
  \begin{itemize}
  \item Get detailed information required for the SLR (journals or conferences, publishers, year of publication, thematic content, etc.).
  \item Understand and establish the basis of an experiment, theoretical framework or model, etc.
  \end{itemize}\\
 \hline
\end{tabular}
\end{table}

According to the breakdown for each of the RQs defined, the details and the categorization type that will be used to classify the selected papers are shown in Table  \ref{TablaRQ-details-and-categories}. This categorization is defined as open (partial) if it does not cover all the possibilities, and therefore more categories could be added. On the opposite, a closed (complete) classification schema covers the whole set of possibilities for that criterion.

\begin{table}[!ht]
    \begin{tabular}{p{0.6cm}|p{8.5cm}|p{1.0cm}}
    \hline
    \textbf{RQ} & \textbf{Detail} & \textbf{CType} \\ 
    \hline 
   	RQ1 & To establish whether the feature modelling tool was used, we established the domain categories, and each paper was assigned to a category according to the domain where the tool was used.   & Open \\
	\hline
 	RQ2	& To study the evidence about the underlying model that each tool is linked, we examined the information provided by each paper and assigned it to one of the defined categories. & Open	 \\
	\hline
 	RQ3	& To establish whether the feature modelling tool was developed from a need of the researchers or satisfied a deficiency detected by the industry, we created three categories, and each paper was assigned to a category according its origin. & Closed \\
 	\hline
 	RQ4	& To quantify how the results provided were validated, we established seven validation categories, and each paper was assigned to a category according to the type of validation.   & Open\\
    \hline
    \end{tabular}
    \caption{RQ - details and classification type (CType: Open, Closed)}
    \label{TablaRQ-details-and-categories}
\end{table}

\subsection{SMS Tool Support}
\label{SMStoolSupport}
\begin{figure}[!ht]\centering
\includegraphics[width=1.0\textwidth]{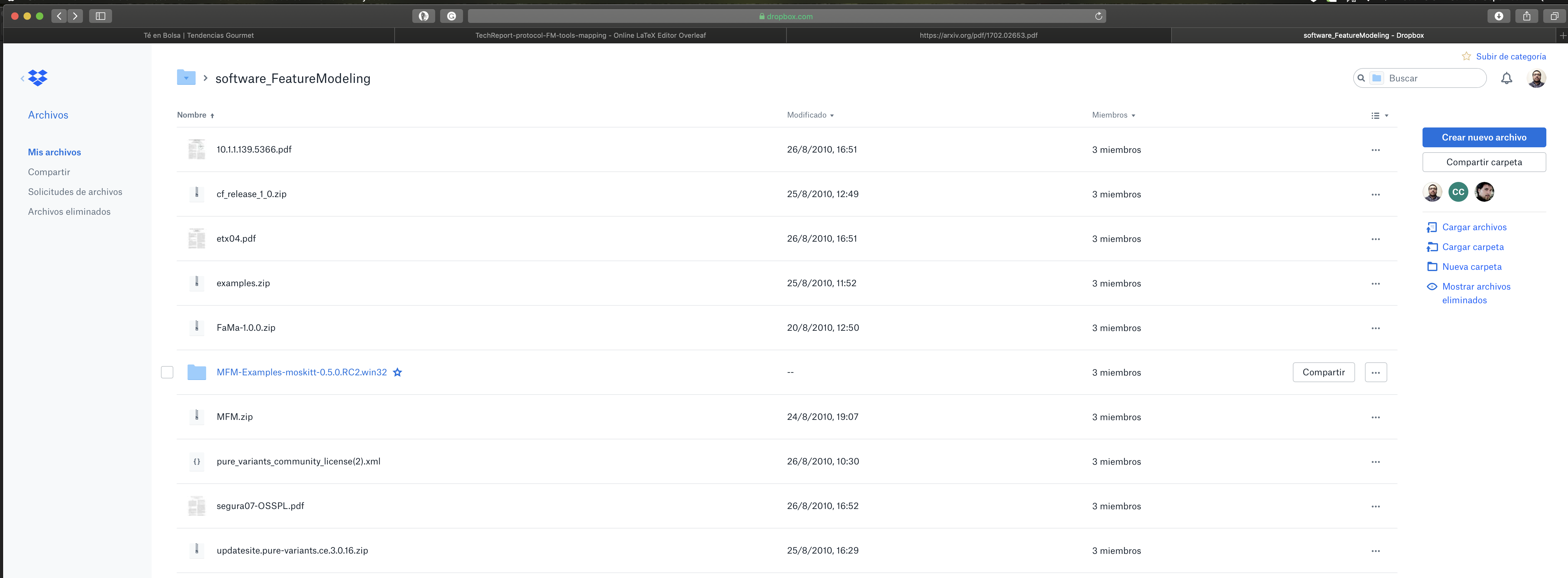}
\caption{Dropbox as a shared repository for resources.} \vspace{-1mm}
\label{fig:dropbox}
\end{figure}

In order to facilitate finding, selecting, documenting and analyzing the information gathered, the following support tools were used: Dropbox\footnote{\url{http://www.dropbox.com}} as a shared repository of resources \cite{drago2012inside}. The details of using this tool are shown in figure \ref{fig:dropbox}.

\begin{figure}[htbp]\centering
\includegraphics[width=1.0\textwidth]{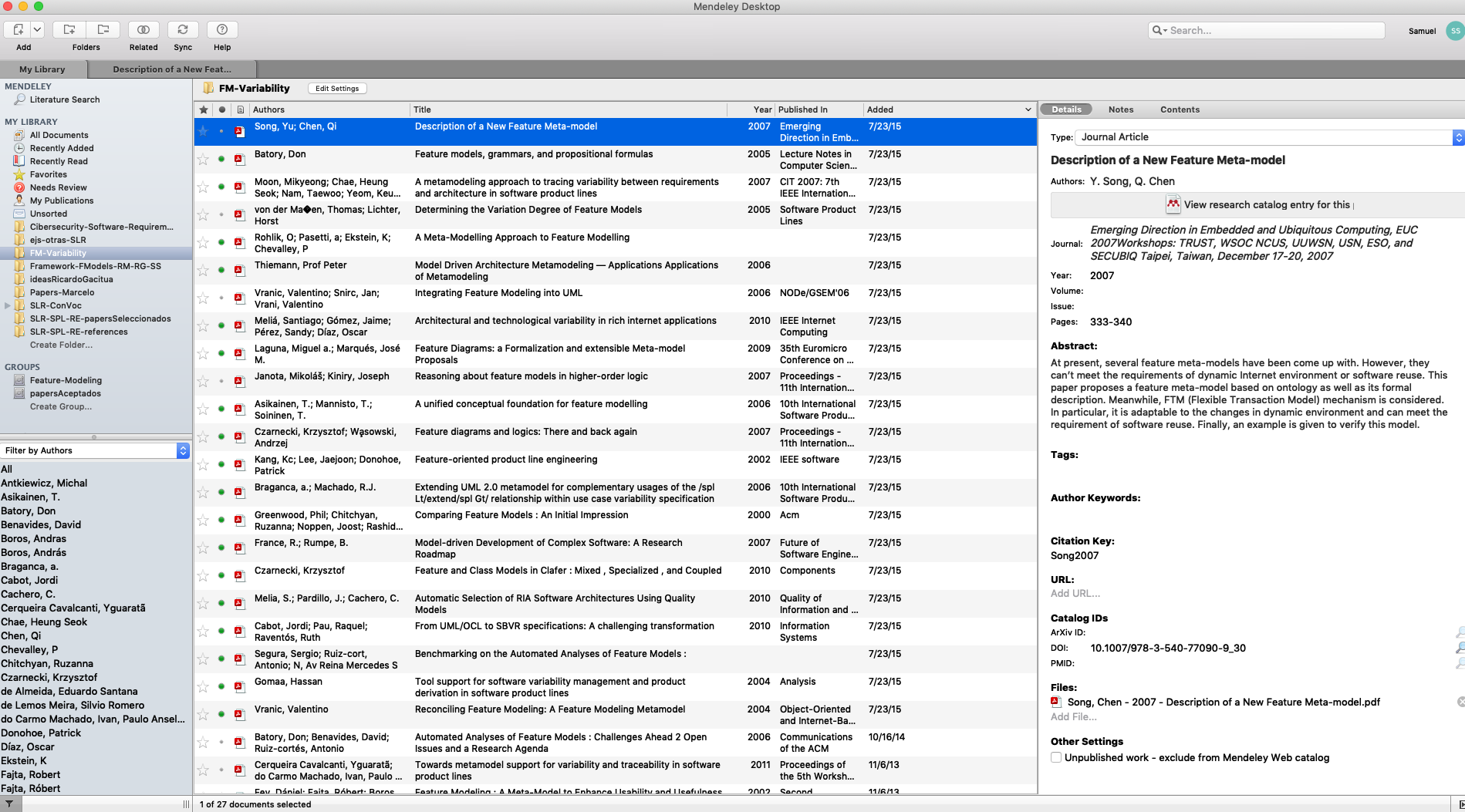}
\caption{Mendeley -desktops version- for storing and reviewing papers.} \vspace{-1mm}
\label{fig:mendeleyDesktop}
\end{figure}

Mendeley\footnote{\url{http://www.mendeley.com}} -Desktop and Web- for storing, reading and annotating reviews for selected papers as well as the automatic creation of the \emph{.bib} files for managing the bibliographic references \cite{vaidhyanathan2012making}. The details of using these tools are shown in figure \ref{fig:mendeleyDesktop}. 

\begin{figure}[!ht]\centering
\includegraphics[width=1.0\textwidth]{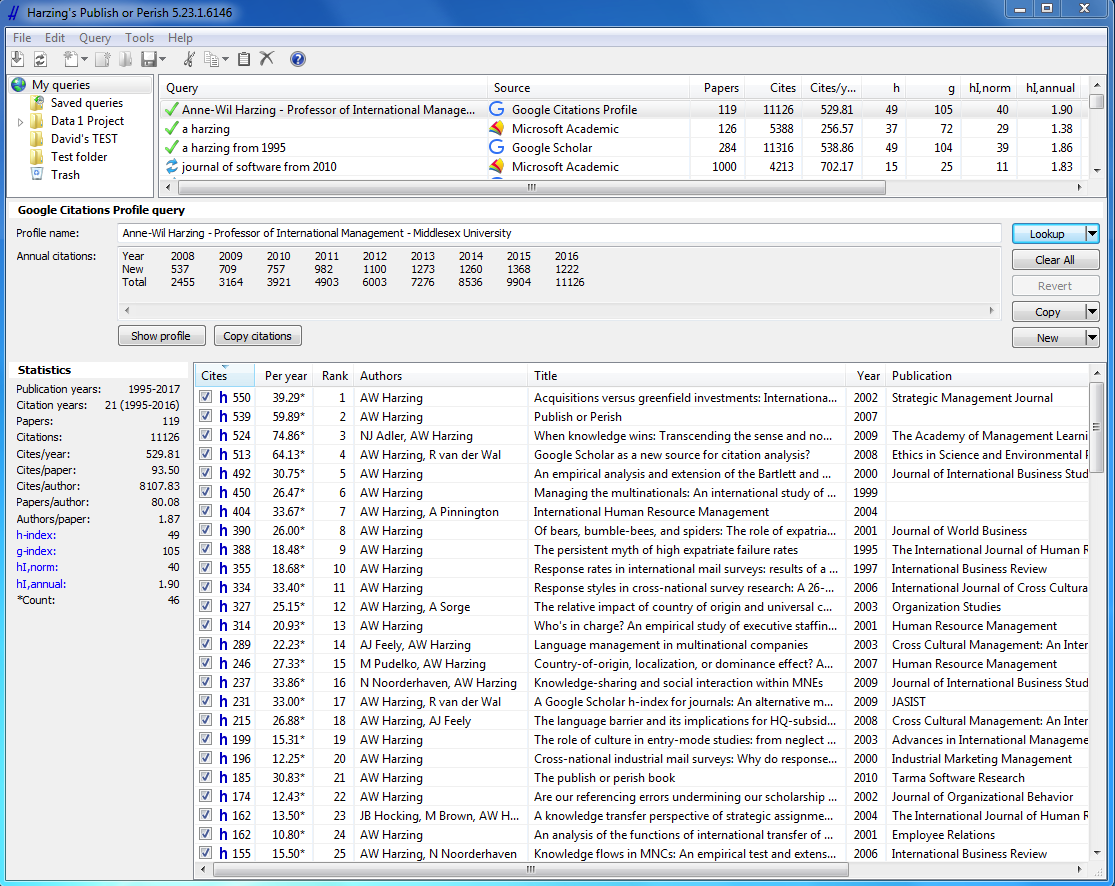}
\caption{Screenshot from website of Publish or Perish tool.} \vspace{-1mm}
\label{fig:pop}
\end{figure}

Publish or Perish\footnote{http://www.harzing.com/resources/publish-or-perish} for initial validation of the search string and automatic spreadsheet creation \cite{harzing2011publish}. The details of using this tool are shown in figure \ref{fig:pop}.

Overleaf\footnote{\url{http://www.overleaf.com}} for editing, managing and controlling the different file versions used to create this paper. The details of using this tool are shown in figure \ref{fig:overleaf}.

\begin{figure}[!ht]\centering
\includegraphics[width=1.0\textwidth]{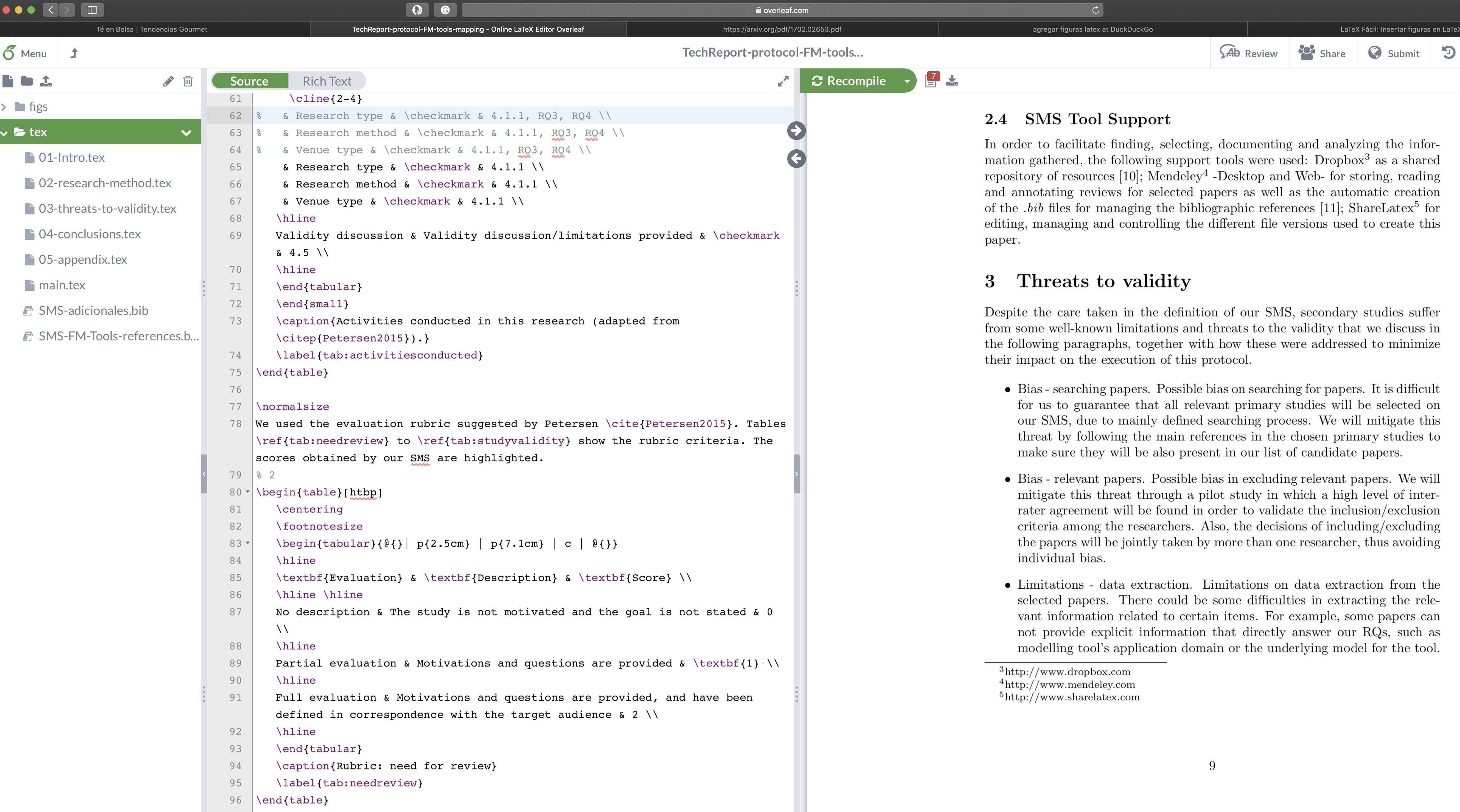}
\caption{Overleaf for writing this report and the paper.} \vspace{-5mm}
\label{fig:overleaf}
\end{figure}
\section{Threats to validity}
\label{Sec:Threats}

Despite the care taken in the definition of our SMS, 
secondary studies suffer from some well-known limitations and threats to the validity that we discuss in the following paragraphs, together with how these were addressed to minimize their impact on the execution of this protocol.



\begin{itemize}
\item Bias - searching papers. 
Possible bias on searching for papers. It is difficult for us to guarantee that all relevant primary studies will be selected on our SMS, due to mainly defined searching process.
We will mitigate this threat by 
following the main references in the chosen primary studies to make sure they will be also present in our list of candidate papers.



\item Bias - relevant papers.
Possible bias in excluding relevant papers. We will mitigate this threat through a pilot study in which a high level of inter-rater agreement will be found in order to validate the inclusion/exclusion criteria among the researchers. Also, the decisions of including/excluding the papers will be jointly taken by more than one researcher, thus avoiding individual bias.


\item Limitations - data extraction.
Limitations on data extraction from the selected papers. There could be some difficulties in extracting the relevant information related to certain items. For example, some papers can not provide explicit information that directly answer our RQs, such as modelling tool’s application domain or the underlying model for the tool.

\item Limitations - searching in data sources.
Limitations of the tools used to conduct searches in the electronic data sources, as already mentioned in Section \ref{ProtocolDefinition} and Table \ref{TablaSearchString}.



We mitigated this threat by talking with experts in SMS and SLR, who gave us feedback, and helped us to validate our defined protocol.





\end{itemize}

Finally, we pretend to check the performance of executing the SMS Protocol according to the rubrics defined by \cite{Petersen2015}. The details are shown (plus the expected results) in the appendix \ref{appendix:Evaluation}.
\section{Conclusions}
\label{Sec:Conclusions}
We have followed the guidelines to plan a SMS according to Petersen \cite{Petersen2015}. As the whole authors adhered to these guidelines to build up the protocol presented in this document, we think that the conducting phase of the  SMS will be repeatable. Finally, the threats to validity have been identified, and mitigated as much as possible.

\section*{Acknowledgements}
Samuel Sepúlveda would like to thank to Dr. Pedro Rossel and Ms.(c) Alonso Bobadilla for their useful advices and technical support.

\bibliographystyle{IEEEtran}
\bibliography{tex/SMS-FM-Tools-references}

\appendix

\section{Evaluation of the SMS process}
\label{appendix:Evaluation}

\normalsize
Here we include a self evaluation for the work that will be done according to \cite{Petersen2015}. Table \ref{tab:activitiesconducted} shows the activities considered for conducting a SMS. We declare using a check-mark (\checkmark) the activities that will be performed.


\footnotesize
\begin{table}[htbp]
   \centering
   \begin{small}
   \footnotesize
   \begin{tabular}{@{}| p{2.4cm} | p{5cm} | c | @{}} 
   \hline
   \textbf{Protocol Phase} & \textbf{Actions} & \textbf{Applied}  \\
   \hline \hline
   Need for map  & Motivate the need and relevance & \checkmark  \\
	&    Define objectives and questions &  \checkmark  \\
	& Consult with target audience to define questions & \tiny \textbullet  \\
   \hline
   Study identification & \multicolumn{2}{|l|}{\textit{Choose search strategy}}  \\
     	\cline{2-3}
	& Snowballing & \tiny \checkmark \\
	& Manual & \tiny \textbullet \\
	& Conduct database search & \checkmark  \\
	 \cline{2-3}
	& \multicolumn{2}{|l|}{\textit{Develop the search}}  \\
	 \cline{2-3}
	& PICOC & \checkmark  \\
	& Consult librarians or experts & \tiny \textbullet  \\
	& Iteratively try to find more relevant papers  & \tiny \textbullet \\
	& Keywords from known papers  & \tiny \textbullet \\
	& Use standards, encyclopedias, and thesaurus  & \tiny \textbullet  \\
	 \cline{2-3}
	& \multicolumn{2}{|l|}{\textit{Evaluate the search} } \\
	 \cline{2-3}
	& Test-set of known papers  & \tiny \textbullet  \\
	& Expert evaluates result  & \tiny \textbullet  \\
	& Search web pages of key authors  & \tiny \textbullet  \\
	& Test--retest  & \tiny \textbullet \\
	 \cline{2-3}
	& \multicolumn{2}{|l|}{\textit{Inclusion and Exclusion criteria}}  \\
	 \cline{2-3}  
	& Identify objective criteria for decision  & \checkmark  \\
	& Add additional reviewer, resolve disagreements between them when needed  & \tiny \textbullet \\
	& Decision rule  & \tiny \textbullet \\
   \hline
   Data extraction and classification & \multicolumn{2}{|l|}{\textit{Extraction process} } \\
   	 \cline{2-3}
   	& Identify objective criteria for decision & \checkmark  \\
	& Obscuring information that could bias & \tiny \textbullet  \\
	& Add additional reviewer, resolve disagreements between them when needed & \tiny \textbullet \\
	& Test--retest & \tiny \textbullet  \\
	 \cline{2-3}
	& \multicolumn{2}{|l|}{\textit{Classification scheme}}\\
	 \cline{2-3}
	& Research type & \checkmark  \\
	& Research method & \checkmark  \\
	& Venue type & \checkmark  \\
   \hline
   Validity discussion & Validity discussion/limitations provided & \checkmark  \\
   \hline
   \end{tabular}
   \end{small}
   \caption{Activities to be conducted in the SMS planned (adapted from \cite{Petersen2015}).}
   \label{tab:activitiesconducted}
\end{table}

\normalsize
We used the evaluation rubric suggested by Petersen \cite{Petersen2015}. Tables \ref{tab:needreview} to \ref{tab:studyvalidity} show the rubric criteria. The scores that will pretend to obtain executing this protocol are highlighted. These scores must will be contrasted with the results at the SMS executing and reporting results.
\begin{table}[htbp]
   \centering
   \footnotesize
   \begin{tabular}{@{}| p{2.5cm} | p{7.1cm} | c | @{}}
   \hline
   \textbf{Evaluation} & \textbf{Description} & \textbf{Score} \\
   \hline \hline
   No description & The study is not motivated and the goal is not stated & 0 \\
   \hline
   Partial evaluation & Motivations and questions are provided & 1 \\
   \hline
   Full evaluation & Motivations and questions are provided, and have been defined in correspondence with the target audience & \textbf{2} \\
   \hline
   \end{tabular}
   \caption{Rubric: need for review}
   \label{tab:needreview}
\end{table}

\begin{table}[htbp]
   \centering
   \footnotesize
   \begin{tabular}{@{}| p{2.5cm} | p{7.1cm} | c | @{}}
   \hline 
   \textbf{Evaluation} & \textbf{Description} & \textbf{Score} \\
   \hline \hline
   No description & Only one type of search has been conducted & 0 \\
   \hline
   Minimal evaluation & Two search strategies have been used & \textbf{1} \\
   \hline
   Full evaluation & All three search strategies have been used & 2 \\
   \hline
   \end{tabular}
   \caption{Rubric: choosing the search strategy}
   \label{tab:choosingstrategy}
\end{table}

\begin{table}[htbp]
   \centering
   \footnotesize
   \begin{tabular}{@{}| p{2.5cm} | p{7.1cm} | c | @{}}
   \hline
   \textbf{Evaluation} & \textbf{Description} & \textbf{Score} \\
   \hline \hline
   No description & No actions have been reported to improve the reliability of the search and inclusion/exclusion criteria & 0 \\
   \hline
   Minimal evaluation & At least one action has been taken to improve the reliability of the search or the reliability of the inclusion/exclusion criteria & 1 \\
   \hline
   Partial evaluation & At least one action has been taken to improve the reliability of the search and the inclusion/exclusion criteria & \textbf{2} \\
   \hline
   Full evaluation & All actions identified have been taken & 3 \\
   \hline
   \end{tabular}
   \caption{Rubric: evaluation of the search}
   \label{tab:evaluationsearch}
\end{table}

\begin{table}[htbp]
   \centering
   \footnotesize
   \begin{tabular}{@{}| p{2.5cm} | p{7.1cm} | c | @{}}
   \hline
   \textbf{Evaluation} & \textbf{Description} & \textbf{Score} \\
   \hline \hline
   No description & No actions have been reported to improve on the extraction process or
enable comparability between studies through the use of existing classifications & 0 \\
   \hline
   Minimal evaluation & At least one action has been taken to increase the reliability of the
extraction process & 1 \\
   \hline
   Partial evaluation & At least one action has been taken to increase the reliability of the extraction process, and research type and method have been classified & \textbf{2} \\
   \hline
   Full evaluation & All actions identified have been taken & 3 \\
   \hline
   \end{tabular}
   \caption{Rubric: extraction and classification}
   \label{tab:extractionclassification}
\end{table}

\begin{table}[!ht]
   \centering
   \footnotesize
   \begin{tabular}{@{}| p{2.5cm} | p{7.1cm} | c | @{}}
   \hline 
   \textbf{Evaluation} & \textbf{Description} & \textbf{Score} \\
   \hline \hline
   No description & No threats or limitations are described & 0 \\
   \hline
   Full evaluation & Threats and limitations are described & \textbf{1} \\ 
   \hline
   \end{tabular}
   \caption{Rubric: study validity}
   \label{tab:studyvalidity}
\end{table}

\end{document}